\documentclass[
    reprint,
    amsmath,amssymb,
    aps,
    prl,
    superscriptaddress,
    floatfix
]{revtex4-2}

\usepackage{bm}        
\usepackage{braket}    
\usepackage{dcolumn}   
\usepackage{graphicx}
\graphicspath{{images/}}
\usepackage{CJKutf8}
\usepackage[percent]{overpic}
\usepackage{hyperref}
\hypersetup{
    colorlinks=true,
    linkcolor=blue,
    citecolor=blue,
    urlcolor=blue
}
\usepackage[capitalize]{cleveref} 
\crefname{equation}{Eq.}{Eqs.}
\Crefname{equation}{Eq.}{Eqs.}
\creflabelformat{equation}{#2(#1)#3}
\crefname{figure}{Fig.}{Figs.}
\Crefname{figure}{Fig.}{Figs.}
\crefname{section}{Sec.}{Secs.}
\Crefname{section}{Sec.}{Secs.}
\crefname{appendix}{App.}{Apps.}
\Crefname{appendix}{App.}{Apps.}

\def\se{\mathcal{S}_{s}}

\def\i{\textnormal{i}}
\def\d{\textnormal{d}}

\def\SO{\textnormal{SO}}

\def\BCS{\text{BCS}}

\def\nn{\nonumber\\&}

\def\c{c^\dagger}

\def\upa{\uparrow}
\def\doa{\downarrow}

\begin{document}

	\title{Measurement-enhanced entanglement in a monitored superconducting chain}

    \author{Rui-Jing Guo \begin{CJK}{UTF8}{gbsn}(郭睿婧)\end{CJK}}
\affiliation{Center for Neutron Science and Technology, Guangdong Provincial Key Laboratory of Magnetoelectric Physics and Devices, School of Physics, Sun Yat-sen University, Guangzhou 510275, China}

\author{Ji-Yao Chen}
\email{chenjiy3@mail.sysu.edu.cn}
\affiliation{Center for Neutron Science and Technology, Guangdong Provincial Key Laboratory of Magnetoelectric Physics and Devices, School of Physics, Sun Yat-sen University, Guangzhou 510275, China}

\author{Zhi-Yuan Wei \begin{CJK}{UTF8}{gbsn}(魏志远)\end{CJK}}
\email{zywei@umd.edu}
\thanks{Current Address: University of Maryland, College Park}
\affiliation{
Max-Planck-Institut f{\"{u}}r Quantenoptik, Hans-Kopfermann-Str. 1, 85748 Garching, Germany
}
    \begin{abstract}
    A common view in monitored quantum dynamics is that local measurements suppress entanglement growth. We show that this intuition can fail in a one-dimensional spinful fermionic chain governed by a BCS Hamiltonian with pairing strength $\Delta$ and subject to continuous, on-site, spin-resolved charge measurements at rate $\gamma$. Using free-fermion simulations and quasiparticle analysis, we show that pairing suppresses entanglement growth, while measurements suppress pairing. Their competition yields \textit{measurement-enhanced entanglement}: for $\Delta>0$, the steady-state entanglement $\se$ increases with $\gamma$ over a finite interval $0<\gamma<\gamma_{\rm peak}$. This occurs because stronger measurements suppress pairing correlations, which would otherwise suppress entanglement growth. Using a nonlinear sigma-model calculation and free-fermion simulations, we provide evidence that for $\Delta>0$ and small but finite $\gamma$, the steady-state entanglement scales as $\se(L)\sim \ln^2 L$. This implies that, in this setting, measurement-enhanced entanglement does not persist in the thermodynamic limit.
\end{abstract}

\maketitle

\textit{Introduction.}---The study of entanglement dynamics in monitored quantum many-body systems has emerged as a new frontier in nonequilibrium physics, bridging condensed matter physics and quantum information science. Arising from the competition between unitary dynamics, which enhance entanglement, and measurements, which suppress it, measurement-induced phase transitions (MIPT) have been identified in a wide variety of systems, including quantum circuits~\cite{Fisher_2023_random,Jian_PhysRevB.106.134206,jian_measurementinduced_2023,Choi_2020,Agrawal_2022_PRX,Bao_2020_mar,Jian_2020_PRB,LYD_2019,Gullans_2020_PRX,Gullans_2020_PRL,Iaconis_2020,lavasani_2021,Sang_PRR,Adam_circuit_2021,Block_int_circuit,wei2025measurement,tikhanovskaya2023universality,lavasani_2021_Nature,Entanglement_PRX_2023}, free ~\cite{fava_adam_U1,Coppola_2022_growth,Adam_PhysRevX.9.031009,Schomerus_PRB_2019,travaglino_quench_2025,swann2023spacetimepictureentanglementgeneration,chan_prb_2019,Cao_2019_SP,Ladewig_prb_2022,Carollo_2022_prb_ALBA,XheK_PRB_2023jul,Marco_PRB_2022jul,Marco_PhysRevB.105.L241114,Van_prl_MAR_2021,Graham_2023_SCIPOST,Poboiko_2D_PhysRevLett.132.110403,poboiko_PhysRevX.13.041046,Poboiko_2D_PhysRevLett.132.110403,Fava_PhysRevX.13.041045,cx_2020_emergent, Zhu_2024,Oshima_2025, Adam_major_defect, DiFresco2026entanglementgrowth, Gal_2023_va} and  interacting~\cite{Poboiko_int,Goto_2020,Fuji_2020,Doggen_2022_PRR,Doggen_2023,foster2025interMIPT} fermionic systems, spin systems~\cite{gal2025_Ising,Sara2023monitoredIsing,Marco2024MEEIsing}, bosonic systems~\cite{Tang_2020,pokharel2025}, disordered systems~\cite{Szyniszewski_PRB_2023_OCT}, and Sachdev-Ye-Kitaev-type models~\cite{JIAN_prl_127_2021,Altland_PhysRevResearch.4.L022066}. Moreover, MIPT have been experimentally realized in trapped ions~\cite{noel_measurement-induced_2022} and superconducting processors~\cite{Koh_2023,wu2025, hoke_measurement-induced_2023,Mart_n_V_zquez_2024}.

A standard expectation in monitored quantum dynamics is that local measurements suppress entanglement growth. This expectation follows from the property that local projective measurements disentangle individual sites from the remainder of the system, and local operations and classical communication (LOCC) cannot generate entanglement~\cite{Nielsen2000}. While entanglement transitions and volume-law states can be generated by measurements alone when using non-commuting multi-site entangling measurements~\cite{matteo2026prxmipt}, single-site non-entangling monitoring is expected to only hinder entanglement. Nevertheless, there is no general proof that the steady-state entanglement entropy \(\se\) must decrease monotonically with the monitoring rate. This raises an intriguing question: \textit{Can single-site non-entangling measurements enhance entanglement growth in monitored dynamics?}

\begin{figure}[b]
    \centering
    \includegraphics[width=0.98\linewidth]{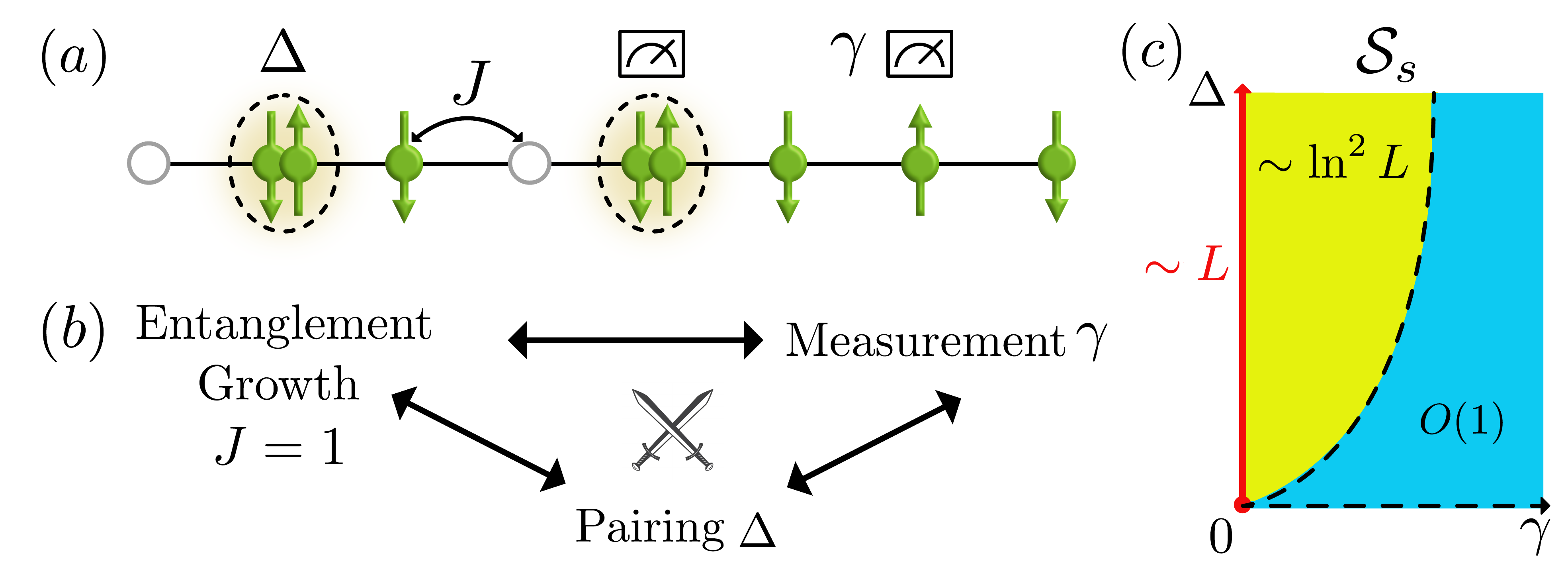}
\caption{\label{fig:schematic}
\textbf{Monitored BCS dynamics}. (a) We consider a 1D system of spinful fermions of $L$ sites, evolving under the BCS Hamiltonian [\cref{eq:hamil}], with hopping amplitude $J=1$. The BCS pairing is illustrated by the dashed circles, with amplitude $\Delta$. The system is subjected to continuous spin-resolved charge measurements of strength $\gamma$. (b) A three-way competition among entanglement growth (induced by fermion hopping), measurement, and pairing. (c) The schematic phase diagram for the scaling of the steady-state entanglement $\se$ as a function of the pairing amplitude $\Delta$ and measurement strength $\gamma$.
}
\end{figure}

In this Letter, we answer the above question in the affirmative by studying the entanglement dynamics of a one-dimensional (1D) chain of spinful fermions of length $L$, evolving under a Bardeen–Cooper–Schrieffer (BCS) Hamiltonian with pairing strength $\Delta$ and continuous spin-resolved charge monitoring at rate $\gamma$ [cf.~\cref{fig:schematic}(a)]. Using both free-fermion simulations and analytical quasiparticle analysis, we show that in the absence of monitoring ($\gamma=0$), BCS pairing suppresses entanglement growth, while the steady state still exhibits volume-law entanglement, $\mathcal{S}_s(L)=c_{\Delta}\cdot L$, with $c_{\Delta}$ decreasing monotonically with $\Delta$. Moreover, measurements suppress the amplitude of the BCS pairing correlations during the dynamics. Taken together, these effects place the entanglement dynamics in a three-way competition among entanglement growth driven by fermion hopping, measurements, and BCS pairing [cf.~\cref{fig:schematic}(b)]. We show that this competition gives rise to the phenomenon of \textit{measurement-enhanced entanglement} (MEE): for $\Delta>0$, the steady-state entanglement $\se$ increases with $\gamma$ over a finite interval $0<\gamma<\gamma_{\rm peak}$. We attribute this phenomenon to the fact that measurements can influence entanglement growth in two distinct ways: they can either directly suppress it, or suppress the BCS pairing, which in turn enhances the entanglement [cf.~\cref{fig:schematic}(b)]. This picture further leads to the prediction that $\gamma_{\rm peak}$ should increase with $\Delta$, which we confirm numerically.

Finally, we study the scaling of the steady-state entanglement. It was previously shown that, for 1D free-fermion monitored dynamics with an arbitrary $U(1)$-conserving Hamiltonian part~\cite{Coppola_2022_growth,fava_adam_U1,Poboiko_int,poboiko_PhysRevX.13.041046}, the steady-state entanglement obeys an area law for any $\gamma>0$. Using a nonlinear sigma-model (NLSM) calculation [see the companion paper~\cite{guo2026NLSM}] and free-fermion simulations, we provide evidence that for $\Delta>0$ and small but finite $\gamma$, the steady-state entanglement scales as $\se(L)\sim \ln^2 L$ [cf.~\cref{fig:schematic}(c)]. This result implies that MEE disappears in the thermodynamic limit, i.e., $\gamma_{\rm peak}\to 0$ as $L\to\infty$, consistent with the finite-size trend of $\gamma_{\rm peak}$ in our numerics.

\textit{Setup.}---We consider a 1D chain of length $L$ of spinful fermions described by the BCS Hamiltonian [cf.~\cref{fig:schematic}(a)]
\begin{align} \label{eq:hamil}
    H_{\BCS} = &-J \sum_{j=1}^L \sum_{\sigma \in (\upa,\doa)} \left(\c_{j,\sigma}c_{j+1,\sigma} + \text{H.c.}\right) \nn- \Delta \sum_{j=1}^L \left(\c_{j,\upa} \c_{j,\doa} + \text{H.c.}\right),
\end{align}
where $\c_{j,\sigma}$ creates a fermion at site $j$ with spin $\sigma \in (\upa,\doa)$. We impose periodic boundary conditions, with $c_{L+1,\sigma} \equiv c_{1,\sigma}$. Throughout this paper, we set the hopping amplitude to $J=1$ as the unit, while $\Delta$ denotes the pairing amplitude. The local charge occupation is given by $n_{j,\sigma}\equiv c^{\dagger}_{j,\sigma} c_{j,\sigma}$, and we continuously monitor this observable at each site $j$ with rate $\gamma$.

Specifically, the dynamics is implemented through a Trotterized evolution: at each time step, the system first undergoes unitary evolution under $e^{-\i H_{\BCS} \delta t}$ with $\delta t \ll 1$, and is then followed by stochastic projective measurements of $n_{j,\sigma}$, performed independently on each site with probability $p = \gamma \delta t$. We choose $\delta t = 0.01$ for all numerical results presented in this paper. During this evolution, the state remains Gaussian, so the entanglement dynamics can be computed exactly using the covariance-matrix formalism, i.e., via free-fermion simulations~\cite{supp,Coppola_2022_growth}. We take the N\'eel state $\ket{\psi_0}=\ket{\upa_1 \doa_2 \dots}$ as the initial state, and show in the Supplemental Material (SM) that the key phenomena are qualitatively similar for the vacuum initial state~\cite{supp}. We are interested in the dynamics of the trajectory-averaged half-chain bipartite entanglement entropy 
\begin{equation} \label{ent_S_eq}
{\cal S}(t)=\langle-{\rm Tr}[\rho_A(t)\ln \rho_A(t)] \rangle_{\rm trajectory},
\end{equation}
where $\rho_A(t)$ is the half-chain reduced density matrix for a single realization,  and we use $\sim 10^3$ trajectories for all numerical results presented in this work. We also denote the steady-state entanglement as $\se \equiv \lim_{t\to \infty} {\cal S}(t)$.

Compared with the setup of 1D monitored hopping fermions~\cite{fava_adam_U1,poboiko_PhysRevX.13.041046,Coppola_2022_growth}, the main new ingredient in our setup is the BCS pairing term [cf.~\cref{eq:hamil}]. This term not only breaks the $U(1)$ symmetry of the Hamiltonian, but also elevates the conventional MIPT setting—governed by a two-party competition between entanglement growth and measurements—into a three-way interplay that additionally involves BCS pairing [cf.~\cref{fig:schematic}(b)]. Accordingly, in the following we first examine the relation of pairing to entanglement growth and to measurements separately, before turning to the full setting.

\begin{figure}[b]
    \centering
    \includegraphics[width=0.98\linewidth]{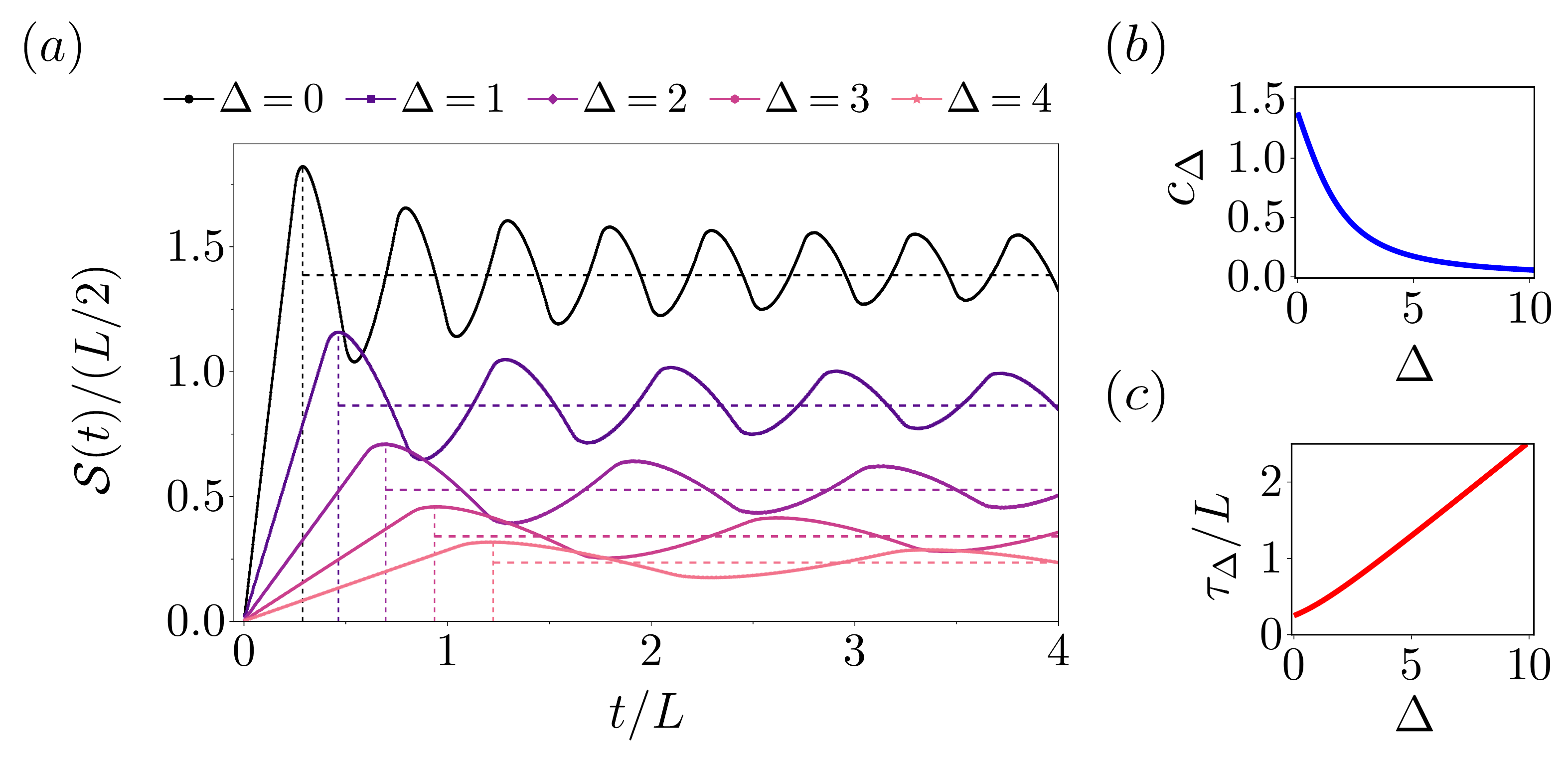}
\caption{\label{fig:unitary}
\textbf{Unmonitored evolution} ($\gamma = 0$).
(a) Time evolution of the entanglement ${\cal S}(t)$ for system size $L=500$ and various pairing strengths $\Delta$. The solid curves show the free-fermion simulations, while the dashed lines indicate the GGE predictions for the steady-state entanglement $\se$ (horizontal) and the entanglement-growth timescale $\tau_{\Delta}$ (vertical). (b) and (c) show the GGE-predicted entropy density $c_\Delta$ [cf.~\cref{Eq:cd}] and the timescale $\tau_{\Delta}$ as functions of $\Delta$. 
}
\end{figure}

\textit{Pairing suppresses entanglement growth.}---Considering the unmonitored evolution under $H_{\rm BCS}$ ($\gamma=0$), \cref{fig:unitary}(a) shows the entanglement ${\cal S}(t)$ for system size $L=500$. We observe a linear growth of ${\cal S}(t)$ toward a volume-law steady-state value $\se$, followed by oscillations around it. Importantly, increasing $\Delta$ systematically reduces ${\cal S}(t)$ both during the growth regime and in its steady-state value, and also extends the timescale $\tau_{\Delta}$ of the linear entanglement growth.

\begin{figure*}[t]
\centering
\includegraphics[width=0.88\linewidth]{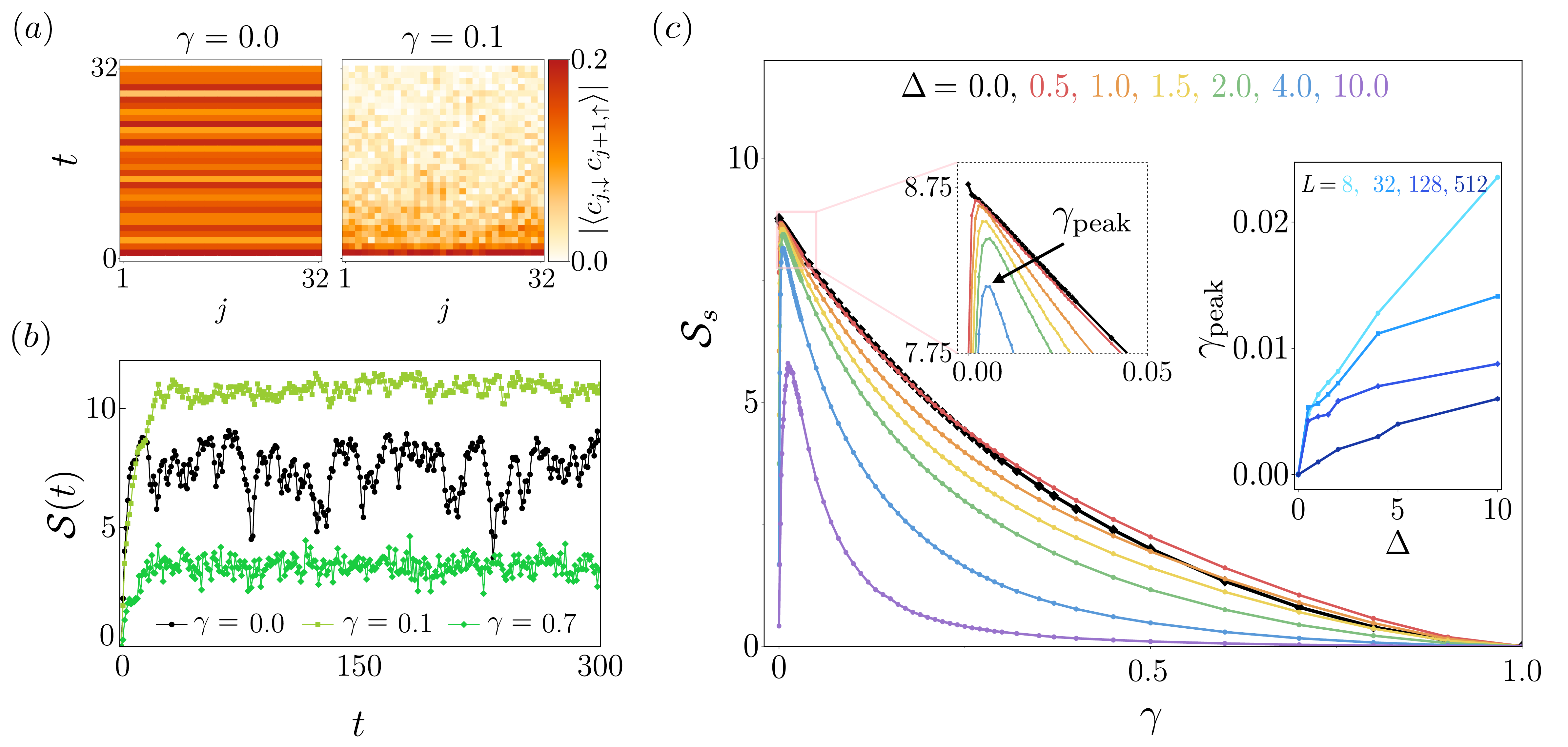}
\caption{\label{fig:entropy_data}
\textbf{Measurement-enhanced entanglement}. System size $L=32$ here.
(a) Evolution of the pairing-correlation amplitude $|\langle c_{j,\downarrow} c_{j+1,\uparrow} \rangle(t)|$ for unmonitored and monitored evolution ($\gamma=0.0,0.1$).  (b) Evolution of the entanglement $\mathcal{S}(t)$ at pairing strength $\Delta = 2.0$, for various measurement strengths $\gamma = 0, 0.1, 0.7$.
(c) The steady-state entanglement $\se$ as a function of $\gamma$, for various values of $\Delta$. The left inset shows a magnified view of the boxed region. Here, $\gamma_{\rm peak}$ denotes the measurement strength at which $\se$ reaches its maximum value. The right inset shows  $\gamma_{\rm peak}$ as a function of $\Delta$ for various fixed $L$.
}
\end{figure*}
To probe the behavior of entanglement dynamics toward the thermodynamic limit ($L\to\infty$), we use the Generalized Gibbs Ensemble (GGE) approach~\cite{BCS_PhysRev.108.1175,bogoljubov_new_1958,Chang_PhysRevB.15.2651} to analytically predict the scaling of $\se$ as well as the timescale $\tau_{\Delta}$, see SM for details~\cite{supp}. GGE predicts a volume-law scaling $\se(L/2) = c_{\Delta}\cdot (L/2)$ for subsystem size $L/2$, with the entropy density $c_{\Delta}$ as:
\begin{align}\label{Eq:cd}
    c_{\Delta} = 2 \int_{-\pi}^{\pi} \frac{\d k}{2\pi} \mathcal{F}\left(\frac{1}{2} + \frac{\Delta}{2E_k}\right).
\end{align}
Here, $\mathcal{F}(y)=-y\ln y-(1-y)\ln (1-y)$ is the binary entropy function, and $E_k =\sqrt{4J^2 {\cos}^2 k + \Delta ^2}$ is the BCS quasiparticle energy. Equation \eqref{Eq:cd} shows that $c_{\Delta}$ monotonically decreases with $\Delta$, as numerically shown in \cref{fig:unitary}(b). Moreover, the timescale for the entanglement growth can be characterized by the group velocity $v_k$ of the quasi-particle, with $|v_k| = |\partial E_k/\partial k|= 2J^2 |\sin(2k)|/{E_k}$. This leads to a timescale $\tau_\Delta\equiv L/(2v_{\max})$ with $v_{\max} = \max_k |v_k|$. Pairing also slows down quasiparticle propagation, as shown in \cref{fig:unitary}(c).

We mark the $\se$ and $\tau_{\Delta}$ predicted by the GGE as the horizontal and vertical lines in \cref{fig:unitary}(a), finding good agreement with the free-fermion simulations. This confirms the GGE prediction that pairing suppresses entanglement growth by limiting the propagation of quasiparticles generated during the quench dynamics.

\textit{Measurements suppress pairing correlations.}--- Evolution under the BCS Hamiltonian generates pairing correlations $\langle c_{i\downarrow} c_{j\uparrow} \rangle$. Because spin-resolved local charge measurements project sites onto definite occupation numbers, they dephase particle-number superpositions and suppress these correlations. As shown in \cref{fig:entropy_data}(a), the nearest-neighbor pairing amplitude $|\langle c_{j\downarrow} c_{j+1\uparrow} \rangle(t)|$ remains finite without measurements ($\gamma=0$), but is rapidly suppressed under monitoring. In the SM, we show that the on-site pairing correlation vanishes in the unmonitored steady state, $\langle c_{j\downarrow} c_{j\uparrow} \rangle_s=0$, and that measurements suppress steady-state intersite pairing correlations across all distances~\cite{supp}.

\textit{Measurement-enhanced entanglement (MEE).}---
Building on the preceding analysis, we confirm that BCS pairing suppresses entanglement growth [cf.~\cref{fig:unitary}], while measurements in turn suppress the pairing [cf.~\cref{fig:entropy_data}(a)]. Together with the direct suppression of entanglement by measurement, the monitored BCS dynamics therefore exhibit a three-way competition [cf.~\cref{fig:schematic}(b)]. In the following, we investigate the resulting surprising phenomenon of MEE.

\Cref{fig:entropy_data}(b) shows the time evolution of the entanglement ${\cal S}(t)$ for several measurement strengths, $\gamma=0.0, 0.1, 0.7$, at fixed pairing strength  $\Delta=2.0$ and system size $L=32$. In all cases, ${\cal S}(t)$ exhibits an initial growth toward a steady-state value, followed by fluctuations around it. These fluctuations are larger than those in \cref{fig:unitary}(a), due to stronger finite-size effects. Remarkably, increasing the measurement rate from $\gamma=0$ to $\gamma=0.1$ enhances the steady-state entanglement $\se$. For a larger measurement rate $\gamma=0.7$, the $\se$ is instead reduced.

We extract the steady-state entanglement, $\se$, by taking a late-time temporal average along each trajectory [$150 \le t \le 300$ for $L=32$, cf.~\cref{fig:entropy_data}(b)], followed by an ensemble average. The results are shown in \cref{fig:entropy_data}(c). In the absence of pairing ($\Delta=0$), $\se$ decreases monotonically with $\gamma$, consistent with previous studies~\cite{fava_adam_U1,Poboiko_int,poboiko_PhysRevX.13.041046,poboiko_PhysRevX.13.041046,Coppola_2022_growth}. In contrast, for $\Delta>0$, $\se$ depends nonmonotonically on $\gamma$: it first increases, reaches a maximum at $\gamma=\gamma_{\rm peak}$, and then decreases for $\gamma>\gamma_{\rm peak}$.

We attribute the behavior shown in \cref{fig:entropy_data}(b,c) to the three-way competition among entanglement growth, measurement, and pairing [cf.~\cref{fig:schematic}(b)]. Specifically, increasing the measurement strength affects entanglement growth through two competing channels: (i) direct suppression of entanglement growth, and (ii) suppression of pairing correlations, which themselves hinder entanglement growth. For small measurement strength, $0<\gamma<\gamma_{\rm peak}$, the channel (ii) dominates because substantial pairing correlations are present, leading to a net enhancement of entanglement growth as $\gamma$ increases.

The above two-channel mechanism operates whenever pairing is present, consistent with \cref{fig:entropy_data}(c), where $\gamma_{\rm peak}>0$ for all values of $\Delta$ considered. Furthermore, $\gamma_{\rm peak}$ characterizes the balance point between channels (i) and (ii). We therefore expect that increasing $\Delta$ broadens the regime in which channel (ii) dominates, thereby increasing $\gamma_{\rm peak}$. This is numerically confirmed in the inset of \cref{fig:entropy_data}(c), where $\gamma_{\rm peak}$ increases with the pairing strength $\Delta$ for various fixed system sizes $L$. Moreover, for fixed $\Delta$, $\gamma_{\rm peak}$ decreases with increasing $L$. This raises the question of whether MEE survives in the thermodynamic limit, and the answer will be closely tied to the system’s steady-state entanglement phase diagram.

\begin{figure}[b]
    \centering    \includegraphics[width=0.98\linewidth]{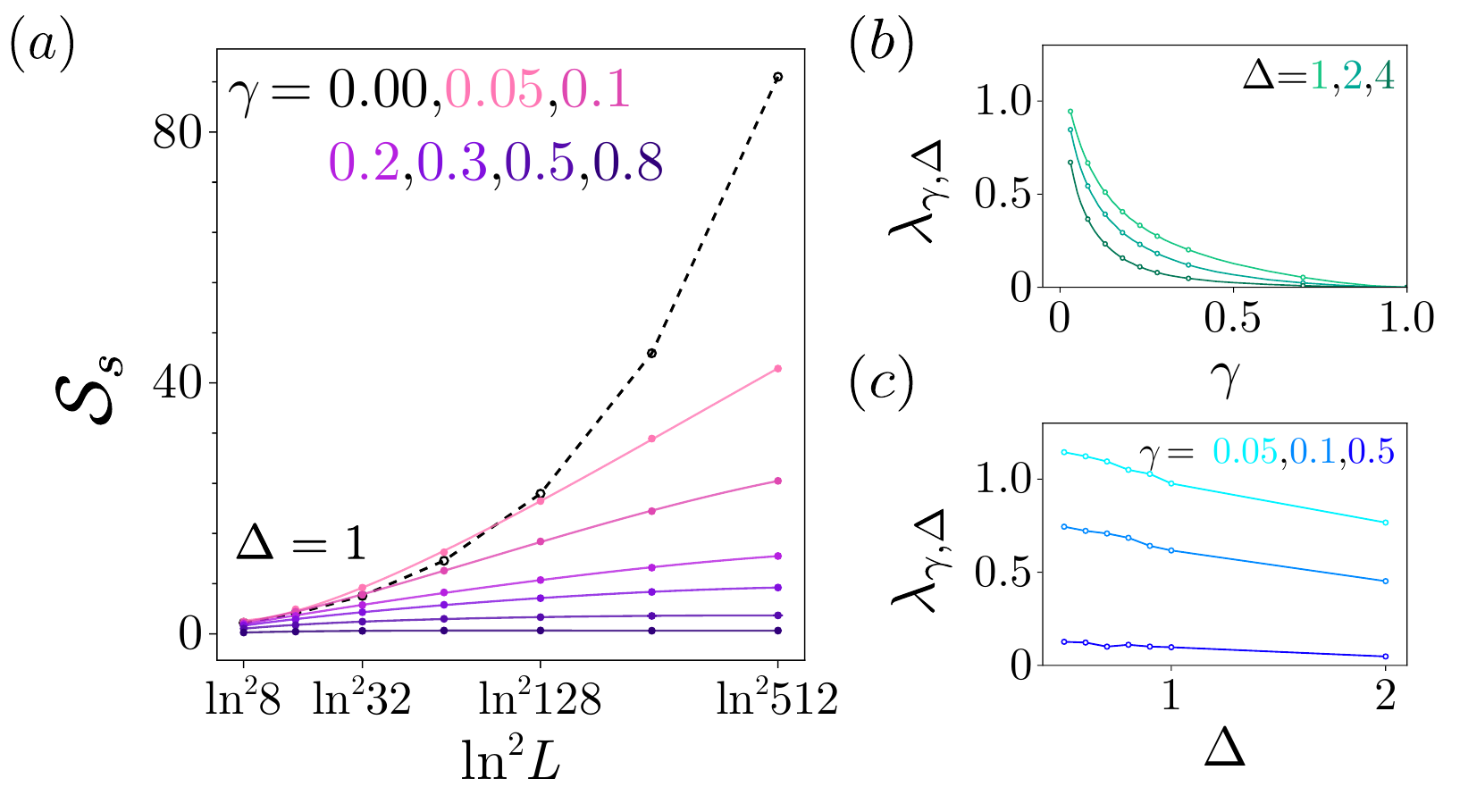}
     \caption{\label{fig:phase}
          \textbf{Steady-state entanglement scaling}. (a) Steady-state entanglement $\se$ versus $\ln^2 L$ at
        pairing strength $\Delta = 1$ for various measurement rates $\gamma$. The black dashed
        line represents the unmonitored case ($\gamma = 0$). Colored symbols show the monitored data, and solid lines represent linear fits to $\se = \lambda_{\gamma,\Delta}\ln^2 L + c$. (b) Fitted coefficient
        $\lambda_{\gamma,\Delta}$ versus $\gamma$ for $\Delta = 1, 2, 4$. (c) Fitted coefficient
        $\lambda_{\gamma,\Delta}$ versus $\Delta$ for $\gamma = 0.05, 0.1, 0.5$.
          }
\end{figure}

\textit{Entanglement phase diagram [cf.~\cref{fig:schematic}(c)].}--- The scaling of the steady-state entanglement $\se$ is a central object of interest in monitored dynamics. Existing results for monitored hopping fermions with $\Delta=0$~\cite{fava_adam_U1,Poboiko_int,poboiko_PhysRevX.13.041046,poboiko_PhysRevX.13.041046,Coppola_2022_growth}, together with our analysis of the unitary limit $\gamma=0$ [cf.~\cref{fig:unitary}], establish that $\se$ exhibits volume-law scaling for ($\gamma=0, \Delta\ge 0$), but reduces to an area law for ($\gamma>0, \Delta=0$). In the strong-monitoring limit $\gamma\gg 1$, the system is strongly projected toward product states, and $\se$ is expected to obey an area law. The central remaining question is whether pairing $\Delta>0$ can stabilize entanglement scaling stronger than an area law under small but finite measurement strength.

To investigate the asymptotic behavior of $\se$, we employ a low-energy effective field theory within the Keldysh formalism~\cite{kamenev_keldysh_2009}. Using the replica trick, the trajectory-averaged steady-state entanglement is related to the free-energy cost of twist defects~\cite{Cardy_2007,Pasquale_Calabrese_2004,Calabrese_2009} in a replicated Keldysh field theory. At long wavelengths, the monitored BCS dynamics is described by an $\SO(R)$ nonlinear sigma model (NLSM) for the remaining gapless modes~\cite{poboiko_PhysRevX.13.041046,Poboiko_int,Poboiko_2D_PhysRevLett.132.110403,fava_adam_U1}. In the rare-measurement limit ($\gamma\ll J, \Delta$), a one-loop RG analysis shows that the effective inverse stiffness increases with the RG scale $\ell = \ln(r/l_0)$, i.e., $g^{-1}(\ell)\sim \ell$. Integrating the twist-defect energy density over length scale $L$ yields the super-logarithmic scaling: 
\begin{align}\label{s_scale}
    \se(L)\sim \int^{\ln L}_0{g^{-1}(\ell)}\, d\ell \sim \ln^2 L.
\end{align}
Conversely, in the strong-monitoring regime ($\gamma \gg J, \Delta$), frequent measurements induce quantum Zeno localization, and $\se$ is expected to follow an area law $\se = O(1)$. In the field-theoretic description, this crossover lies beyond the one-loop perturbative regime, where the proliferation of topological defects~\cite{fava_adam_U1,Fu_topology_2012} provides a plausible mechanism for truncating the correlation length. Detailed derivations for the field theory are presented in the companion paper~\cite{guo2026NLSM}.

In \cref{fig:phase}(a), we plot the free-fermion simulation results for the steady-state entanglement $\se$ as a function of $\ln^2 L$ at fixed pairing strength $\Delta=1.0$. In the absence of measurements ($\gamma = 0$), the $\se$ curve gradually bends upward with increasing $L$, indicating volume-law scaling. By contrast, for small but finite $\gamma > 0$, $\se$ depends almost linearly on $\ln^2 L$, providing evidence that $\se$ follows the scaling $\se \sim \ln^2 L$ in the presence of measurements [cf.~\cref{s_scale}]. At larger measurement rates, such as $\gamma=0.8$, the curves are almost flat, suggesting that $\se$ returns to area law scaling in this case. For $\gamma > 0$, we fit the data to the form $\se = \lambda_{\gamma,\Delta}\ln^2 L + c$ and use the effective finite-size fit slope $\lambda_{\gamma,\Delta}$ to characterize how the super-logarithmic scaling depends on the system parameters. At fixed $\Delta$, $\lambda_{\gamma,\Delta}$ decreases monotonically with increasing $\gamma$ [\cref{fig:phase}(b)], indicating that stronger measurements suppress the $\sim\ln^2 L$ growth of $\se$ and drive the system toward the area-law regime. At fixed $\gamma$, $\lambda_{\gamma,\Delta}$ decreases with increasing $\Delta$ but remains finite [\cref{fig:phase}(c)], indicating that stronger pairing suppresses entanglement growth, consistent with the behavior observed in the unmonitored evolution [cf.~\cref{fig:unitary}]. Despite this suppression, we expect the $\se=O(\ln^2 L)$ phase to persist for $\gamma\ll1$ at any finite $\Delta>0$, as predicted by our analytical study based on the NLSM~\cite{guo2026NLSM} and reminiscent of the corresponding behavior in the unmonitored case [cf.~\cref{Eq:cd}]. Overall, the numerical and analytical results provide complementary evidence for the phase diagram shown in \cref{fig:schematic}(c).

The phase diagram in \cref{fig:schematic}(c) implies that the MEE observed in \cref{fig:entropy_data} does not persist in the thermodynamic limit for the 1D monitored superconducting chain considered here. This is because $\se$ scales asymptotically as a volume law at $\gamma=0$, but as at most a super-logarithmic law for arbitrarily small $\gamma>0$. Thus one can always find a sufficiently large system size $L$ such that $\gamma_{\rm peak} \to 0$. This is consistent with the numerical results in the inset of \cref{fig:entropy_data}(c), where $\gamma_{\rm peak}$ decreases with increasing system size. We also remark that $\gamma_{\rm peak}$ remains practically appreciable up to $L=128$, suggesting that the MEE phenomenon can still manifest at experimentally relevant system sizes in moderate-scale quantum devices.

We remark that while $\se \sim \ln^2 L$ also appears in the monitored dynamics of topological $p$-wave Majorana fermions~\cite{Fava_PhysRevX.13.041045}, the monitored $s$-wave superconducting chain studied here exhibits the same super-logarithmic scaling in a topologically trivial system. Our results therefore provide a complementary and distinct example in which breaking $U(1)$ symmetry allows 1D monitored free-fermion dynamics to sustain a beyond-area-law scaling of $\se$ even at small but finite measurement strength.

\textit{Outlook.}---As we have summarized our results in the introduction, here we discuss a few promising directions:

(i) \textbf{MEE in other systems.} Our proposed two-channel mechanism suggests that MEE can arise in a wide range of systems exhibiting a three-way competition [cf.~\cref{fig:schematic}(b)]. It would be interesting to design other setups that display MEE~\cite{Marco2024MEEIsing}, particularly by incorporating unitary disentangling processes into monitored dynamics. Some examples of such unitary disentangling strategies have already been explored in the context of quantum circuits~\cite{morral_2025_disentangling}. (ii) \textbf{MEE in the thermodynamic limit.} As discussed in the main text, a necessary condition for MEE to persist in the thermodynamic limit is the existence of a stable volume-law phase of steady-state entanglement at small but finite measurement strength. However, existing evidence suggests that free-fermion systems under measurement can exhibit only super-logarithmic or area-law scaling~\cite{poboiko_PhysRevX.13.041046,Poboiko_2D_PhysRevLett.132.110403,fava_adam_U1,foster2025interMIPT}. It is therefore an interesting open question whether MEE can persist in other platforms, such as quantum circuits or measurement-only models~\cite{matteo2026prxmipt,morral_2025_disentangling}, where volume-law entanglement survives at finite measurement rates.

\textit{Acknowledgement.}--
RJG and JYC are supported by National Natural Science Foundation of China (Grants No. 12447107, No. 12304186), Guangdong Basic and Applied Basic Research Foundation (Grant No. 2024A1515013065), and Quantum Science and Technology - National Science and Technology Major Project (Grant No. 2021ZD0302100).

\bibliography{ref,library}

\end{document}


\crefname{equation}{Eq.}{Eqs.}
\crefname{figure}{Fig.}{Fig.}
\crefname{appendix}{Appendix}{Appendix}
\renewcommand{\thefigure}{S\arabic{figure}}
\renewcommand{\theequation}{S\arabic{equation}}
\renewcommand{\bibnumfmt}[1]{[S#1]}
\renewcommand{\citenumfont}[1]{S#1}
\renewcommand{\thesection}{S\arabic{section}}

\title{
Supplemental Material to ``Measurement-enhanced entanglement in a monitored superconducting chain''
}

    \author{Rui-Jing Guo \begin{CJK}{UTF8}{gbsn}(郭睿婧)\end{CJK}}
\affiliation{Center for Neutron Science and Technology, Guangdong Provincial Key Laboratory of Magnetoelectric Physics and Devices, School of Physics, Sun Yat-sen University, Guangzhou 510275, China}

\author{Ji-Yao Chen}
\email{chenjiy3@mail.sysu.edu.cn}
\affiliation{Center for Neutron Science and Technology, Guangdong Provincial Key Laboratory of Magnetoelectric Physics and Devices, School of Physics, Sun Yat-sen University, Guangzhou 510275, China}

\author{Zhi-Yuan Wei \begin{CJK}{UTF8}{gbsn}(魏志远)\end{CJK}}
\email{zywei@umd.edu}
\thanks{Current Address: University of Maryland, College Park}
\affiliation{
Max-Planck-Institut f{\"{u}}r Quantenoptik, Hans-Kopfermann-Str. 1, 85748 Garching, Germany
}   

\maketitle

\creflabelformat{equation}{#2(#1)#3}

\def\I{\mathbb{I}}
\def\Z{\mathbb{Z}}
\def\R{\mathbb{R}}
\def\L{\mathcal{L}}
\def\K{\mathcal{K}}
\def\B{\mathcal{B}}
\def\G{\mathcal{G}}
\def\D{\mathcal{D}}
\def\T{\mathcal{T}}
\def\O{\mathcal{O}}
\def\mw{\mathcal{W}}
\def\se{\mathcal{S}_s}
\def\hg{\hat{\mathcal{G}}}
\def\hr{\hat{\mathcal{R}}}
\def\hx{\hat{\mathcal{X}}}
\def\hs{\hat{\Sigma}}
\def\hq{\hat{Q}}
\def\hv{\hat{V}}
\def\hw{\hat{W}}
\def\hm{\hat{M}}
\def\hz{\hat{Z}}
\def\x{\boldsymbol{x}}
\def\k{\boldsymbol{k}}
\def\i{\mathrm{i}}
\def\d{\mathrm{d}}
\def\SU{\mathrm{SU}}
\def\SO{\mathrm{SO}}
\def\Sp{\mathrm{Sp}}
\def\tr{\mathrm{tr}}
\def\Tr{\mathrm{Tr}}
\def\U{\mathrm{U}}
\def\USp{\mathrm{USp}}
\def\eff{\text{eff}}
\def\reg{(\text{reg})}
\def\kin{\text{kin}}
\def\mea{\text{mea}}
\def\all{\text{all}}
\def\BCS{\text{BCS}}
\def\repl{\text{repl}}
\def\diag{\text{diag}}
\def\mass{\text{mass}}
\def\Nambu{\text{Nambu}}
\def\massive{\text{massive}}
\def\massless{\text{massless}}
\def\mani{\mathrm{USp(4R)}/\mathrm{U}(2R)}
\def\rl{\right.\nonumber\\&\left.}
\def\nn{\nonumber\\&}
\def\n{\nonumber\\}
\def\en{\nonumber\\=&}
\def\c{c^\dagger}
\def\p{\psi}
\def\bp{\bar{\psi}}
\def\P{\Psi}
\def\BP{\bar{\Psi}}
\def\upa{\uparrow}
\def\doa{\downarrow}

In this Supplemental Material, we provide details of the analytical derivations and numerical methods for the one-dimensional monitored BCS model studied in the main text.

\section{Gaussian State Simulation of Monitored BCS Dynamics}\label{App:Gamma}

\subsection{Simulating Monitored Dynamics Starting from the N\'eel Product State}\label{App:neel}
We consider a one-dimensional chain of spinful fermions with $L$ sites. The operator $c^{\dagger}_{l,\sigma}$ creates a fermion at site $l \in \{1,\ldots,L\}$ with spin $\sigma \in \{\uparrow,\downarrow\}$. The system is characterized by fermionic creation and annihilation operators that obey the canonical anticommutation relations (CAR): $\{\c_{i,\alpha}, c_{j,\beta}\} = \delta_{i,\alpha, j,\beta}$ and $\{c_{i,\alpha}, c_{j,\beta}\} = \{\c_{i,\alpha}, \c_{j,\beta}\} = 0$ for $\alpha,\beta\in\{\upa,\doa\}$ at site $i,j\in\{1,\dots,L\}$.

The time evolution proceeds in discrete steps of duration $\delta t$. During each interval, the state is propagated by the unitary operator $U(\delta t) = e^{-\i H_\BCS\delta t}$ [cf.~ Eq. (1) in the main text], evolving as:
\begin{align}
    \ket{\psi(t+\delta t)} = U(\delta t)\ket{\psi(t)}.
\end{align}
For the results in the main text, we take the N\'eel state as the initial state:
\begin{align}\label{eq:Neelproduct}
	\ket{\psi(0)}=\prod_{l=0}^{L/2-1}\c_{2l+1,\uparrow}\c_{2l+2,\downarrow}\ket{0}=\ket{\uparrow_1,\downarrow_2,\uparrow_3,\downarrow_4,\dots,\upa_{L-1},\doa_L},
\end{align}
where spin-up fermions occupy the odd-numbered sites and spin-down fermions occupy the even-numbered sites.

Following the unitary evolution $U(\delta t)$, stochastic projective measurements are applied independently at each site with probability $p=\gamma \delta t$ in the single-site basis ${\ket{0},\ket{\uparrow},\ket{\downarrow},\ket{\uparrow\downarrow}}$, where $\gamma$ denotes the measurement rate.
We implement the local charge measurement via sequential projective measurements of $n_{l,\uparrow}$ and $n_{l,\downarrow}$. Given that the occupation number operators for opposite spin species commute, $[n_{l,\uparrow},n_{l,\downarrow}] = 0$, this sequential protocol is mathematically equivalent to a direct orthogonal projection onto the single-site basis $\{\ket{0},\ket{\uparrow},\ket{\downarrow},\ket{\uparrow\downarrow}\}$.

For a specific site $l$, the four possible projectors are:
\begin{align}\label{eq:projmea}
	&\hat{P}_{0,l} = (1-n_{l,\uparrow})(1-n_{l,\downarrow})=\ket{0_l}\bra{0_l} 
    \nn\hat{P}_{\uparrow,l} = n_{l,\uparrow}(1-n_{l,\downarrow})=\ket{\uparrow_l}\bra{\uparrow_l}
    \nn\hat{P}_{\downarrow,l} = (1-n_{l,\uparrow})n_{l,\downarrow}=\ket{\downarrow_l}\bra{\downarrow_l} \nn\hat{P}_{\uparrow \downarrow,l} = n_{l,\uparrow}n_{l,\downarrow}=\ket{\uparrow_l,\downarrow_l} \bra{\uparrow_l,\downarrow_l}		
\end{align}
The occupation number operator $n_{l,\sigma}=\c_{l,\sigma}c_{l,\sigma}$ has eigenvalues of $0$ and $1$ at site $l$ with spin flavor $\sigma$.

 To facilitate the simulation of Gaussian dynamics, we introduce the Nambu spinor $\vec{C}$ of dimension $4L$, defined as:
\begin{align}\label{eq:Nambu_spinor}
    \vec{C} = \begin{pmatrix}
        c_{x} \\ \c_{x}
    \end{pmatrix}, \quad 
    \vec{C}^\dagger = \begin{pmatrix}
        \c_{x} & c_{x}
    \end{pmatrix},
\end{align}
where the collective index $x = (1\upa, \dots, L\upa, 1\doa, \dots, L\doa)$ enumerates the degrees of freedom across the one-dimensional chain, explicitly distinguishing between lattice sites and spin flavors.

A Fermionic Gaussian state $|\psi(t)\rangle$ on a $L$-particle system is fully characterized by the covariance matrix: 
\begin{align}\label{gammat_eq}
	\Gamma(t)= \bra{\psi(t)}\vec{C}\vec{C}^\dagger\ket{\psi(t)} =\begin{pmatrix}
		\langle c_x\c_x\rangle_t&\langle c_xc_x\rangle_t\\
		\langle \c_x\c_x\rangle_t&\langle \c_x c_x\rangle_t
	\end{pmatrix}=\begin{pmatrix}
		\Gamma^{11}&\Gamma^{12}\\\Gamma^{21}&\Gamma^{22}
	\end{pmatrix}_t,
\end{align}
with the block structure explicitly given by:
\begin{align}
    \Gamma^{ab}=\begin{pmatrix}
		\Gamma^{ab}_{i,\uparrow,j\uparrow}&\Gamma^{ab}_{i,\uparrow,j\downarrow}\\
		\Gamma^{ab}_{i,\downarrow,j\uparrow}&\Gamma^{ab}_{i,\downarrow,j\downarrow}
	\end{pmatrix},
\end{align}
where $a,b\in(1,2)$, and $j,l\in(1,\dots,L)$ denote the site indices. The matrix elements obey the following fermionic symmetry constraints:
\begin{align}\label{sym_constrain}
	&\Gamma^{11}_{i,\alpha,j,\beta} = \delta_{i,\alpha,j,\beta}-\Gamma^{22}_{j,\beta,i,\alpha},
	\\&\Gamma^{12}_{i,\alpha,j,\beta}=-\Gamma^{12}_{j,\beta,i,\alpha}=-(\Gamma^{21}_{i,\alpha,j,\beta})^*=(\Gamma^{21}_{j,\beta,i,\alpha})^*,
\end{align}
where $\alpha,\beta\in(\upa,\doa)$.

Numerically, to simulate the local charge measurement, we generate a single uniform random number $r_l \in [0,1]$ for each site index $l$ at every time step. If $r_l < p$, the site $l$ is selected for measurement. To implement the projection onto the four-dimensional local charge basis while preserving the Gaussianity of the state, we sequentially perform the projective measurements of the spin-up ($n_{l,\uparrow}$) and spin-down ($n_{l,\downarrow}$) number operators on that specific site, evaluating the respective Born probabilities at each step.

During the sequential projection on the selected site $l$, each spin species $\sigma$ is processed one after the other. For a given spin $\sigma$, following Born's rule, the outcome depends on the pre-measurement occupation probability $\langle n_{l,\sigma} \rangle = \Gamma^{22}_{l,\sigma,l,\sigma}$. We generate a fresh, independent uniform random number $x_{l,\sigma} \in [0,1]$ to determine the collapse. The local state is projected onto the occupied sector ($1$) with probability $\langle n_{l,\sigma} \rangle$, and the empty sector ($0$) with probability $1-\langle n_{l,\sigma} \rangle$. Since the initial state is Gaussian, the Hamiltonian is quadratic, and the measurements preserve Gaussianity~\cite{Coppola_2022_growth}, the system remains Gaussian throughout the monitored evolution. This allows us to track the dynamics simply by updating the covariance matrix using Wick's theorem. In the following, we discuss the generic update rules for the blocks $\Gamma^{22}$ and $\Gamma^{12}$ under a single projection of species $\sigma$, while the remaining blocks, $\Gamma^{11}$ and $\Gamma^{21}$, are fixed by the symmetry relations in \cref{sym_constrain}.

\paragraph{Case 1: Occupied Outcome.---}
If the outcome is $1$, which occurs when $x_{l,\sigma} \leq \langle n_{l,\sigma} \rangle$, the state is projected as:
\begin{align} \label{eq:occup}
	|\psi\rangle\rightarrow\cfrac{n_{l,\sigma}|\psi\rangle}{\sqrt{\langle n_{l,\sigma} \rangle}}.
\end{align}
The covariance matrix block $\Gamma^{22}$ updates to:
\begin{align} \label{eq:G22_proj}
	\Gamma_{i,\alpha,j,\beta}^{22}\rightarrow\cfrac{\langle\psi|n_{l,\sigma}\c_{i,\alpha}c_{j,\beta}n_{l,\sigma}|\psi\rangle}{\langle n_{l,\sigma} \rangle},
\end{align}
Applying anti-commutation relations and Wick's theorem yields the explicit update rule:
\begin{align} \label{eq:G22_proj_yes}
	\Gamma_{i,\alpha,j,\beta}^{22}\rightarrow\Gamma_{i,\alpha,j,\beta}^{22}+\delta_{l,\sigma,i,\alpha}\delta_{l,\sigma,j,\beta}-\cfrac{\Gamma_{i,\alpha,l,\sigma}^{22}\Gamma_{l,\sigma,j,\beta}^{22}+\Gamma_{j,\beta,l,\sigma}^{12}\left(\Gamma_{l,\sigma,i,\alpha}^{12}\right)^{*}}{\langle n_{l,\sigma} \rangle}.
\end{align}
Similarly, the off-diagonal block $\Gamma^{12}$ updates as:
\begin{align} \label{eq:G12_proj}
	\Gamma_{i,\alpha,j,\beta}^{12}\rightarrow\cfrac{\langle\psi|n_{l,\sigma}c_{i,\alpha}c_{j,\beta}n_{l,\sigma}|\psi\rangle}{\langle n_{l,\sigma} \rangle}=\Gamma_{i,\alpha,j,\beta}^{12}+\cfrac{\Gamma_{j,\beta,l,\sigma}^{12}\Gamma_{l,\sigma,i,\alpha}^{22}-\Gamma_{i,\alpha,l,\sigma}^{12}\Gamma_{l,\sigma,j,\beta}^{22}}{\langle n_{l,\sigma} \rangle}.
\end{align}

\paragraph{Case 2: Empty Outcome.---}
If the outcome is $0$, which occurs when $x_{l,\sigma} > \langle n_{l,\sigma} \rangle$, the state is projected as:
\begin{align} \label{eq:nooccup}
	|\psi'\rangle=\cfrac{\left(1-n_{l,\sigma}\right)|\psi\rangle}{\sqrt{1-\langle n_{l,\sigma} \rangle}}.
\end{align}
The $\Gamma^{22}$ block updates according to:
\begin{align} \label{eq:2Gamma22}
	\Gamma_{i,\alpha,j,\beta}^{22}&\rightarrow\cfrac{\langle\psi|\left(1-n_{l,\sigma}\right)\c_{i,\alpha}c_{j,\beta}\left(1-n_{l,\sigma}\right)|\psi\rangle}{1-\langle n_{l,\sigma} \rangle}\nonumber \\
    &=-\delta_{l,\sigma,i,\alpha}\delta_{l,\sigma,j,\beta}+\Gamma_{i,\alpha,j,\beta}^{22}+\cfrac{\Gamma_{j,\beta,l,\sigma}^{12}\left(\Gamma_{l,\sigma,i,\alpha}^{12}\right)^{*}+\left(\delta_{l,\sigma,i,\alpha}-\Gamma_{i,\alpha,l,\sigma}^{22}\right)\left(\delta_{l,\sigma,j,\beta}-\Gamma_{l,\sigma,j,\beta}^{22}\right)}{1-\langle n_{l,\sigma} \rangle}.
\end{align}
The pairing correlation block $\Gamma^{12}$ updates as:
\begin{align} \label{eq:2Gamma12}
	\Gamma_{i,\alpha,j,\beta}^{12}&\rightarrow\cfrac{\langle\psi|\left(1-n_{l,\sigma}\right)c_{i,\alpha}c_{j,\beta}\left(1-n_{l,\sigma}\right)|\psi\rangle}{1-\langle n_{l,\sigma} \rangle} \nonumber \\
    &=\Gamma_{i,\alpha,j,\beta}^{12}+\cfrac{\Gamma_{j,\beta,l,\sigma}^{12}\left(\delta_{l,\sigma,i,\alpha}-\Gamma_{l,\sigma,i,\alpha}^{22}\right)-\Gamma_{i,\alpha,l,\sigma}^{12}\left(\delta_{l,\sigma,j,\beta}-\Gamma_{l,\sigma,j,\beta}^{22}\right)}{1-\langle n_{l,\sigma} \rangle}.
\end{align}

\subsection{Evolution starting from vacuum state}
\begin{figure}[t]
    \centering
    \includegraphics[width=0.76\linewidth]{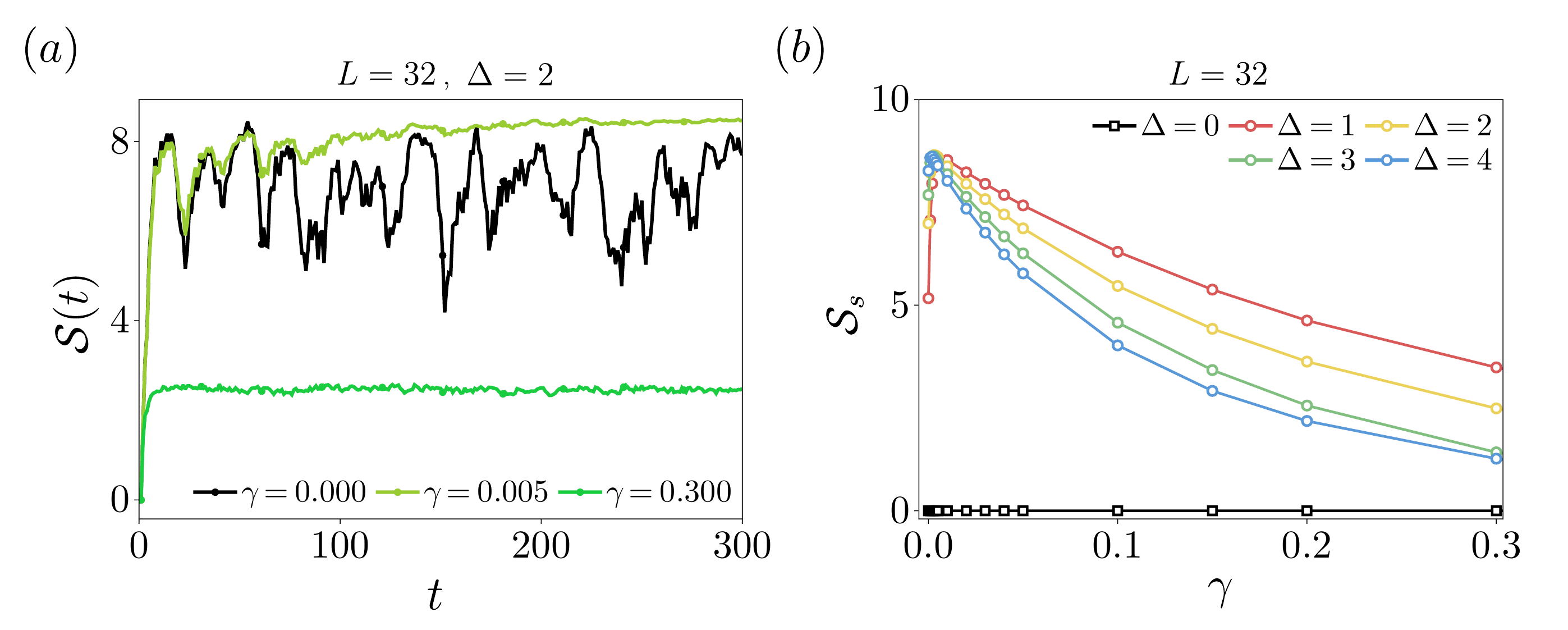}
    \caption{\label{fig:vac}
    Monitored BCS dynamics starting from the vacuum state $|\psi_{\rm vac}(0)\rangle = |0,\ldots,0\rangle$ for system size $L=32$. (a) Time evolution of the entanglement ${\cal S}(t)$ for fixed pairing strength $\Delta=2$ and several measurement rates $\gamma$. (b) Steady-state entanglement $\se$ as a function of $\gamma$ for various fixed pairing strengths $\Delta$.}
\end{figure}

Besides the initial N\'eel state studied in the main text, in \cref{fig:vac} we also consider the same monitored BCS dynamics starting from the vacuum state, $|\psi_{\rm vac}(0)\rangle = |0,\ldots,0\rangle$. Similar to Fig.~3(b,c) in the main text, we again observe the same measurement-enhanced entanglement phenomenon, showing that it is not specific to the initial N\'eel state.

\section{Quasiparticle description of unmonitored BCS dynamics}\label{quasi}

We analytically study the unmonitored BCS dynamics ($\gamma=0$) via a quasiparticle description in the Generalized Gibbs Ensemble (GGE) framework~\cite{BCS_PhysRev.108.1175,bogoljubov_new_1958,Chang_PhysRevB.15.2651,schmidt_physics_1997}. The diagonalization of $H_{\rm BCS}$ [Eq. (1) in the main text] proceeds via the discrete Fourier transform:
\begin{equation}
    c_{j,\sigma} = \frac{1}{\sqrt{L}} \sum_k e^{\i kj} c_{k,\sigma}, \quad \c_{j,\sigma} = \frac{1}{\sqrt{L}} \sum_k e^{-\i kj} \c_{k,\sigma}. 
\end{equation}
where $j \in \{1, \dots, L\}$ denotes the spatial site index, and $\sigma \in \{\upa, \doa\}$ denotes the spin flavor. Assuming periodic boundary conditions (PBC), the summation runs over discrete momenta $k = 2\pi n/L$ restricted to the first Brillouin zone $[-\pi, \pi)$, corresponding to integers $n \in \{-L/2, \dots, L/2 - 1\}$.
In momentum space, the Hamiltonian decouples into independent modes:
\begin{equation}
    H_{\rm BCS} = \sum_{k,\sigma} \xi_k \c_{k,\sigma} c_{k,\sigma} - \sum_k \Delta (\c_{k,\upa} \c_{-k,\downarrow} + c_{-k,\downarrow} c_{k,\upa}),
\end{equation}
where $\xi_k = -2J\cos k$ denotes the kinetic energy. Introducing the Nambu spinor basis $\Psi_k = (c_{k,\upa}, \c_{-k,\downarrow})^\mathrm{T}$, the Hamiltonian is written as:
\begin{equation}
    H_{\BCS} = \sum_k \Psi_k^\dagger H_{\text{BdG}}(k) \Psi_k + E_0, \quad H_{\text{BdG}}(k) = \begin{pmatrix} \xi_k & -\Delta \\ -\Delta & -\xi_k \end{pmatrix},
\end{equation}
where $E_0 \equiv \sum_k \xi_k = 0$ due to the tight-banding dispersion. Note that $\Psi_k$ is the momentum-space counterpart to the two-particle sector of the real-space Nambu spinor $\vec{C}$ defined in \cref{eq:Nambu_spinor}, reduced by translation symmetry. 

To diagonalize this quadratic Hamiltonian, we introduce a new set of fermionic operators $\gamma_{k,\sigma}$, which represent the fundamental non-interacting Bogoliubov quasiparticle excitations. The original electron operators are related to these quasiparticles via a unitary Bogoliubov transformation:
\begin{equation}\label{eq:bogo_transform}
    c_{k,\upa} = u_k \gamma_{k,\upa} + v_k \gamma_{-k,\downarrow}^\dagger, \nonumber \qquad \qquad
    \c_{-k,\downarrow} = -v_k \gamma_{k,\upa} + u_k \gamma_{-k,\downarrow}^\dagger,
\end{equation}
By substituting this transformation back into the BdG Hamiltonian, the off-diagonal pairing terms are analytically eliminated. The Hamiltonian can then be mapped onto a system of independent Bogoliubov quasiparticles:
\begin{align}\label{eq:H_diagonal_BdG}
    H_{\text{BCS}} = \sum_{k,\sigma} E_k \gamma_{k,\sigma}^\dagger \gamma_{k,\sigma} + E_{\text{gs}},
\end{align}
characterized by the quasiparticle dispersion relation $E_k = \sqrt{\xi_k^2 + \Delta^2}$, and the ground-state energy of the BCS condensate $E_{\text{gs}} = -\sum_k E_k$, thus the coherence factors are determined by the relations $u_k v_k = \frac{\Delta}{2E_k}$ and $u_k^2 - v_k^2 = \frac{\xi_k}{E_k}$.  In this diagonalized form, it becomes evident that the dynamics of the system are governed by the independent evolution of these free quasiparticle modes.

\subsection{Steady-state pairing correlations}

\subsubsection{Numerical Results}

\begin{figure}[t]
    \centering
    \includegraphics[width=0.54\linewidth]{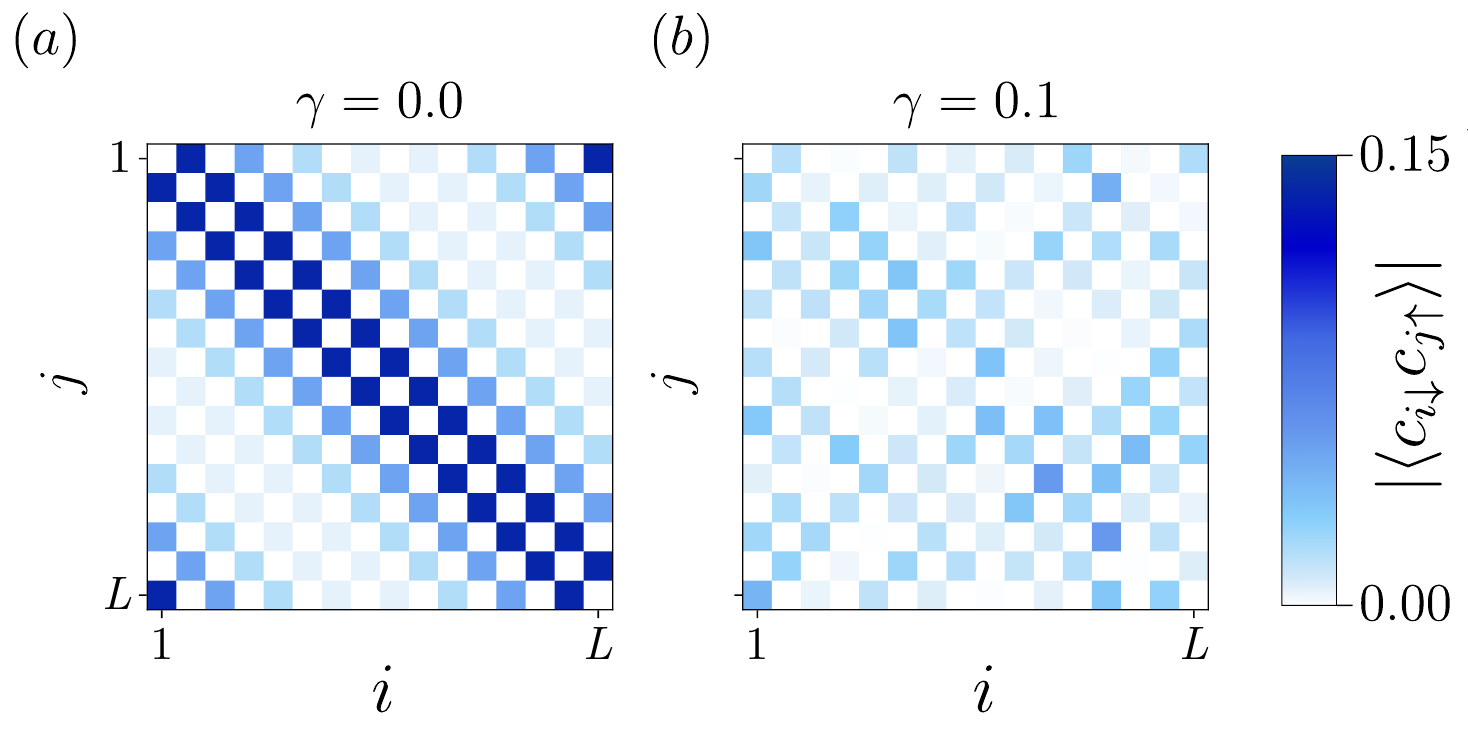}
    \caption{\label{fig:cooper pair}
    Amplitude of the steady-state pairing correlations, $|\langle c_{i,\doa}, c_{j,\upa} \rangle|$. Panels (a) and (b) show the unmonitored ($\gamma = 0$) and monitored ($\gamma = 0.1$) dynamics, respectively. The system size is $L = 32$, and the pairing strength is $\Delta = 1$.}
\end{figure}

Using exact Gaussian-state numerics, we extract the amplitude of the steady-state pairing correlations from the covariance matrix $|\Gamma^{21}|$ [cf.~\cref{gammat_eq}], whose elements are given by $|\Gamma^{21}|_{ij} = |\langle c_{i\doa} c_{j\upa} \rangle|$. The results, shown in \cref{fig:cooper pair}, are obtained for pairing amplitude $\Delta = 1$ and system size $L = 32$.

For the unmonitored dynamics ($\gamma=0$) [cf.~\cref{fig:cooper pair}(a)], $|\Gamma^{21}|$ has off-diagonal bands. Physically, it indicates that nearest-neighbor Cooper pairing correlations exist in the steady state. This pattern arises because the initial N\'eel state corresponds to a high-energy configuration where quasiparticle modes are half-filled ($\langle\c_{k,\sigma}c_{k,\sigma}\rangle=\langle n_{k,\sigma}\rangle=1/2$), thereby blocking the conventional Cooper instability term proportional to $(1-2n_{k,\sigma})$. However, the breaking of single-site translational symmetry in the initial state induces a coupling between momenta $k$ and $k+\pi$. This Umklapp mechanism sustains finite nearest-neighbor pairing correlations $\langle c_{j\doa} c_{j+1\upa} \rangle$. In the following subsections \ref{onsite_apd} and \ref{nearsite_apd}, we further explain this banded structure through an analytical GGE calculation.

When measurements are included ($\gamma = 0.1$) [cf.~\cref{fig:cooper pair}(b)], the steady-state pairing correlations are generally suppressed. As discussed in the main text, this is expected because charge monitoring suppresses the coherence between the zero- and double-fermion occupation sectors required for the existence of pairing correlations. Together with Fig. 3(a) in the main text, these results demonstrate that measurements suppress pairing in the monitored dynamics. 

\subsubsection{GGE prediction of onsite pairing correlations}
\label{onsite_apd}
After quenching from the initial N\'eel state [cf.~\cref{eq:Neelproduct}], the system relaxes to a steady state described by a GGE rather than undergoing conventional thermalization. Since the post-quench Hamiltonian is fully diagonalized in the Bogoliubov quasiparticle basis [cf.~\cref{eq:H_diagonal_BdG}], the commutator of any generic quasiparticle bilinear operator $\gamma_{k,\sigma}^\dagger \gamma_{k'\sigma'}$ with Hamiltonian is given by:
\begin{equation}
    [\gamma_{k,\sigma}^\dagger \gamma_{k'\sigma'}, H_{\text{BCS}}] = (E_{k'} - E_{k})\gamma_{k,\sigma}^\dagger \gamma_{k'\sigma'}
\end{equation}
Consequently, its Heisenberg time evolution is governed by the energy difference:
\begin{equation}
    \gamma_{k,\sigma}^\dagger(t) \gamma_{k'\sigma'}(t) = e^{\i (E_{k} - E_{k'}) t}\gamma_{k,\sigma}^\dagger(0) \gamma_{k'\sigma'}(0).
\end{equation}
When taking the long-time average to obtain the steady-state expectation value, the phase factor $e^{\i(E_{k} - E_{k'}) t}$ oscillates rapidly if $E_{k} \neq E_{k'}$, leading to dynamical dephasing that causes the steady-state expectation value of this term to vanish. Conversely, the phase factor remains unity if and only if $E_{k} = E_{k'}$. Under this condition, the operator needs to commute with the Hamiltonian:
\begin{equation}
    [\gamma_{k,\sigma}^\dagger \gamma_{k'\sigma'}, H_{\text{BCS}}] = 0 \quad (\text{for } E_{k} = E_{k'}).
\end{equation}
Thus, for any bilinear operator $\gamma_{k,\sigma}^\dagger \gamma_{k',\sigma'}$ with $k$ and $k'$ satisfying $E_k = E_{k'}$, the steady-state expectation value is exactly equal to its initial value at $t=0$; that is, such operators are exact integrals of motion of the system.

Therefore, the quasiparticle occupation number operators $n_{k,\sigma}^{(q)} \equiv \gamma_{k,\sigma}^\dagger \gamma_{k,\sigma}$ commute with the Hamiltonian, $[n_{k,\sigma}^{(q)}, H_{\text{BCS}}] = 0$. Consequently, these occupation numbers are integrals of motion. Their steady-state expectation values are strictly determined by their initial values at $t=0$ and remain time-translation invariant throughout the evolution. Explicit evaluation of these conserved quantities in the initial state yields the time-independent quasiparticle populations and pairing correlations:
\begin{equation}\label{eq:n_k_q}
n_{k,\sigma}^{(q)} \equiv \langle \psi(t) | \gamma_{k,\sigma}^\dagger \gamma_{k,\sigma} | \psi(t) \rangle = \langle \psi(0) | \gamma_{k,\sigma}^\dagger \gamma_{k,\sigma} | \psi(0) \rangle = \frac{1}{2},
\end{equation}
\begin{equation}
\kappa_{k,\sigma}^{(q)} \equiv \langle \psi(t) | \gamma_{-k,\downarrow} \gamma_{k,\upa} | \psi(t) \rangle = \langle \psi(0) | \gamma_{-k,\downarrow} \gamma_{k,\upa} | \psi(0) \rangle = 0.
\end{equation}
The vanishing of the quasiparticle pairing correlation $\kappa_k^{(q)}$ is a consequence of the absence of pairing correlations in the initial product state.
It follows that the steady-state pairing correlator in momentum space vanishes identically:
\begin{align}\label{eq:chi_k}
    \chi_k &\equiv \langle c_{k,\upa} c_{-k,\downarrow} \rangle_s = u_k v_k (1 - n_k^{(q)} - n_{-k}^{(q)}) + (u_k^2 \kappa_k^{(q)} - v_k^2 \kappa_k^{(q)*})= u_k v_k (1 - 1) + 0 = 0.
\end{align}
Since $\chi_k$ vanishes uniformly for all discrete momenta $k$ across the Brillouin zone, the on-site pairing correlation in real space evaluates to strictly zero in the steady state for any arbitrary finite system size $L$:
\begin{align}\label{eq:c_jc_j_def}
    \langle c_{j,\downarrow}c_{j,\upa} \rangle_s &= \frac{1}{L}\sum_{k} e^{-ik \cdot 0} (-\chi_k) = \frac{1}{L} \sum_{k} 0 = 0,
\end{align}
which agrees with the exact finite-size numerical results observed in \cref{fig:cooper pair}(a).

\subsubsection{GGE prediction of nearest-neighbor pairing correlations}\label{nearsite_apd}
In contrast to the vanishing on-site pairing implied by $\chi_k = 0$ [cf.~\cref{eq:chi_k}] , we now evaluate the nearest-neighbor pairing correlation $\langle c_{j,\downarrow}c_{j+1,\upa} \rangle_s$ in the steady state. 

The real-space nearest-neighbor pairing correlation is reconstructed via the Fourier expansion:
\begin{align}\label{eq:c_jc_j+1_def}
    \langle c_{j,\downarrow}c_{j+1,\upa} \rangle_s &= \frac{1}{L} \sum_{k,q} e^{\i kj} e^{\i q(j+1)} \langle c_{k,\downarrow} c_{q,\upa} \rangle_s,
\end{align}
where both momenta $k$ and $q$ are summed over the first Brillouin zone $[-\pi,\pi)$ under PBC. The initial N\'eel state breaks the single-site translational symmetry of the 1D lattice. This symmetry breaking leads to a spatial period doubling. Consequently, the fundamental reciprocal lattice vector is halved from $2\pi$ to $\pi$. Because the original translational invariance is reduced to a discrete two-site translation, the crystal momentum is now conserved only modulo $\pi$. This allows the periodic background to provide or absorb a discrete momentum of $\pi$ during scattering events. Such an Umklapp~\cite{GG_1994,giamarchi_quantum_2003} scattering mechanism directly couples the momentum modes $k$ and $k+\pi$. Therefore, the pairing correlation $\langle c_{k,\downarrow} c_{q,\upa} \rangle_s$ can only take non-zero values when the momentum selection rule $q = -k$ (the conventional channel) or $q = -k+\pi$ (the Umklapp channel) is satisfied. According to \cref{eq:chi_k}, the only pairing expectation value that can be nonzero is $\langle c_{k,\downarrow} c_{-k+\pi,\upa} \rangle_s$. We therefore define the Umklapp pairing correlation as follows:
\begin{equation}\label{eq:eta_k_def}
    \eta_k \equiv \langle c_{k,\downarrow} c_{-k+\pi,\upa} \rangle_s.
\end{equation}
To evaluate this quantity, we first recast the fermionic operators into the quasiparticle basis by invoking the Bogoliubov transformation defined in \cref{eq:bogo_transform}. Specifically, the relevant annihilation operators are expressed as $c_{k,\downarrow} = u_k \gamma_{k,\downarrow} - v_k \gamma_{-k,\uparrow}^\dagger$ and $c_{-k+\pi,\uparrow} = u_{-k+\pi} \gamma_{-k+\pi,\uparrow} + v_{-k+\pi} \gamma_{k-\pi,\downarrow}^\dagger$. Utilizing the symmetry relations $u_{k+\pi}=v_k$ and $v_{k+\pi}=u_k$ due to the periodic Brillouin zone, the latter becomes $c_{-k+\pi,\uparrow} = v_k \gamma_{-k+\pi,\uparrow} + u_k \gamma_{k-\pi,\downarrow}^\dagger$.

For simplicity, we define the cross-momentum quasiparticle correlations and their complex conjugates as:
\begin{align}\label{eq:C_k_gamma}
    \mathcal{C}_{k,\sigma} \equiv \langle \gamma_{k,\sigma}^\dagger \gamma_{k+\pi,\sigma} \rangle, \quad \mathcal{C}_{k,\sigma}^* =\langle \gamma_{k+\pi,\sigma}^\dagger \gamma_{k,\sigma} \rangle.
\end{align}
In the Heisenberg picture, the quasiparticle creation and annihilation operators evolve as $\gamma_{k,\sigma}^\dagger(t) = \gamma_{k,\sigma}^\dagger(0) e^{\i E_k t}$ and $\gamma_{k+\pi,\sigma}(t) = \gamma_{k+\pi,\sigma}(0) e^{-\i E_{k+\pi} t}$, respectively. This leads to the time evolution of the cross-momentum operator:
\begin{equation}
    \hat{\mathcal{C}}_{k,\sigma}(t) = \gamma_{k,\sigma}^\dagger(0) \gamma_{k+\pi,\sigma}(0) e^{\i (E_k - E_{k+\pi}) t} = \hat{\mathcal{C}}_{k,\sigma}(0) e^{\i (E_k - E_{k+\pi}) t}.
\end{equation}
Consequently, its expectation value evolves as:
\begin{equation}
    \langle \hat{\mathcal{C}}_{k,\sigma}(t) \rangle = \langle \hat{\mathcal{C}}_{k,\sigma}(0) \rangle e^{\i (E_k - E_{k+\pi}) t}.
\end{equation}
According to the normal-state dispersion relation $\xi_k = -2J\cos k$, the properties of the cosine function imply that a momentum shift of $\pi$ reverses the sign of the kinetic energy:
\begin{equation}
    \xi_{k+\pi} = -2J\cos(k+\pi) = 2J\cos k = -\xi_k.
\end{equation}
This relation reflects the chiral (particle-hole) symmetry inherent to the bipartite lattice. As a result, the quasiparticle excitation energy remains invariant under this momentum shift:
\begin{equation}
    E_{k+\pi} = \sqrt{(-\xi_k)^2 + \Delta^2} = E_k.
\end{equation}
Because $E_k \equiv E_{k+\pi}$, the relative phase factor strictly evaluates to unity, i.e., $e^{\i(E_k - E_{k+\pi})t} \equiv 1$. Therefore, the correlation $\mathcal{C}_{k,\sigma}$ does not dephase to zero over time; instead, it constitutes a conserved quantity within the GGE, maintaining its initial expectation value 
\begin{equation} \label{ck_inv}
\langle \hat{\mathcal{C}}_{k,\sigma}(t) \rangle = \langle \hat{\mathcal{C}}_{k,\sigma}(0) \rangle.	
\end{equation}
 This time-translation invariance implies the commutation relation $[\gamma_{k,\sigma}^\dagger \gamma_{k+\pi,\sigma}, H_{\text{BCS}}] = 0$.

By evaluating these expectation values over the initial N\'eel state [cf.~\cref{eq:Neelproduct}], we obtain the exact GGE steady-state correlations: $\mathcal{C}_{-k\upa} = -u_k v_k =- \frac{\Delta}{2E_k}$ and $\mathcal{C}_{k\downarrow}^* = u_k v_k=\frac{\Delta}{2E_k}$. Noting that cross-spin expectation values (e.g., $\langle \gamma_{k,\downarrow} \gamma_{-k+\pi,\uparrow} \rangle$) naturally vanish due to spin conservation, we substitute these expanded operators into the definition of $\eta_k$ in \cref{eq:eta_k_def}. Expanding the product yields the final expression:
\begin{align}\label{eq:eta_k_final}
    \eta_k = -u_k^2 \mathcal{C}_{k\downarrow}^* - v_k^2 \mathcal{C}_{-k\upa} = -\frac{\Delta \xi_k}{2E_k^2}. 
\end{align}
Substituting \cref{eq:eta_k_final} together with the dispersion relation $\xi_k =-2J\cos k$ into the discrete real-space inverse Fourier transform yields the correlation for an arbitrary finite system size $L$:
\begin{align}\label{eq:real_space_int}
    \langle c_{j,\downarrow}c_{j+1,\upa} \rangle_s &= \frac{1}{L}
      \sum_{k} e^{\i kj} e^{\i (-k+\pi)(j+1)} \eta_k 
    \nn= e^{\i \pi (j+1)} \frac{1}{L} \sum_{k} e^{-\i k} \left( \frac{J \Delta \cos
      k}{E_k^2} \right) 
    \nn= (-1)^{j+1} \frac{1}{L} \sum_{k} (\cos k - \i \sin k) \frac{J \Delta \cos
      k}{E_k^2}.
\end{align}
Because the discrete allowed momenta $k = 2\pi n / L$ under PBC are distributed symmetrically around $k=0$, the discrete sum over the odd-parity term (proportional to $\sin k \cos k$) exactly vanishes. Consequently, the surviving even-parity term, which is proportional to $\cos^2 k$, yields a strictly finite non-zero result for any arbitrary $L$. In the thermodynamic limit ($L \to \infty$), the discrete sum $\frac{1}{L}\sum_k$ converges to the continuous integral $\int_{-\pi}^{\pi}\frac{\d k}{2\pi}$, giving:
\begin{equation}\label{eq:final_staggered}
    \lim_{L \to \infty} \langle c_{j,\downarrow}c_{j+1,\upa}\rangle_s = (-1)^{j+1} \frac{J \Delta}{2\pi} \int_{-\pi}^{\pi} \d k \frac{\cos^2k}{\xi_k^2 + \Delta^2} \neq 0.
\end{equation}
This analytical result confirms the persistence of nearest-neighbor pairing correlations in the steady state of the unmonitored dynamics, consistent with the numerical observations in \cref{fig:cooper pair}(a).

The steady-state pairing correlations and the MEE phenomenon are not exclusive to the N\'eel initial state. For a translationally invariant initial state, such as the vacuum state $|\psi_{\rm vac}(0)\rangle =|0,\ldots,0\rangle$, the Umklapp channel is inactive. The post-quench quasiparticle population deviates from half-filling ($n_k^{(q)}= v_k^2 \neq 1/2$), lifting the Pauli blocking in the conventional momentum channel. This yields a non-zero steady-state pairing expectation value $\chi_k = u_kv_k (1 - 2v_k^2) = \frac{\Delta \xi_k}{2E_k^2}$. Substituting this into the real-space integral gives a spatially uniform nearest-neighbor pairing correlation:
\begin{align} \label{eq:vac_staggered}
    \lim_{L \to \infty} \langle c_{j,\downarrow}c_{j+1,\upa}\rangle_{s, \text{vac}} = \int_{-\pi}^{\pi} \frac{\d k}{2\pi} \frac{J \Delta\cos^2 k}{\xi_k^2 + \Delta^2} \neq 0.
\end{align}
This amplitude matches the staggered correlation of the N\'eel state [cf.~\cref{eq:final_staggered}]. Thus, quenching to the BCS Hamiltonian produces non-zero steady-state pairing correlations whether the initial state breaks translational symmetry or not (via either the Umklapp or the conventional channel). It is also expected that measurements suppress pairing correlations in this case as well. Thus, the two-channel mechanism explained in the main text [cf.~Fig.~1(b) in the main text] also accounts for the measurement-enhanced entanglement observed for the initial vacuum state [cf.~\cref{fig:vac}].

\subsection{Entanglement Entropy}
For a generic fermionic Gaussian state, the von Neumann entanglement entropy $\mathcal{S}$ associated with the bipartition $AA'$ is completely determined by the eigenvalues $\{\lambda_m\}$ of the covariance matrix restricted to subsystem $A$. Specifically, it is given by the extensive summation of the binary entropy over all independent eigenmodes $m$~\cite{Vidal_2003,peschel_calculation_2003}:
\begin{equation} \label{eq:peschel_sum}
    \mathcal{S} = \sum_m \mathcal{F}(\lambda_m), \quad \mathcal{F}(x) \equiv -x \ln x -(1-x) \ln(1-x),
\end{equation}
where the index $m$ enumerates all independent eigenmodes of the system. Since the evolved state remains Gaussian throughout the monitored dynamics, we can use \cref{eq:peschel_sum} to compute the entanglement ${\cal S}(t)$, as well as its steady-state value $\se$. 

Furthermore, due to the breaking of single-site translational symmetry in the initial N\'eel state, Umklapp scattering strictly couples the momentum mode $k$ with $k+\pi$. Because the dynamics also strictly conserve the spin flavor, the independent eigenmodes of the system are formed by the two-dimensional invariant subspaces spanned by $\{ \gamma_{k,\sigma}, \gamma_{k+\pi,\sigma} \}$ for each spin flavor $\sigma \in \{\uparrow, \downarrow\}$. We define $S_{k, k+\pi, \sigma}$ as the entanglement entropy contribution originating from this specific decoupled subspace.

According to the time-translation invariance of $\mathcal{C}_{k,\sigma}$ [cf.~\cref{ck_inv}] and the conserved populations $n_k^{(q)}$ [cf.~\cref{eq:n_k_q}], the steady-state covariance matrix $\Gamma^{(q)}$ decouples into independent $2 \times 2$ blocks. For a given spin $\sigma$, the matrix takes the explicit form:
\begin{equation}\label{eq:Gamma_k_k_pi}
\Gamma_{k, k+\pi, \sigma}^{(q)} = \begin{pmatrix} \langle\gamma_{k,\sigma}^\dagger \gamma_{k,\sigma} \rangle & \langle\gamma_{k,\sigma}^\dagger \gamma_{k+\pi,\sigma} \rangle \\ \langle\gamma_{k+\pi,\sigma}^\dagger \gamma_{k,\sigma} \rangle & \langle\gamma_{k+\pi,\sigma}^\dagger \gamma_{k+\pi,\sigma} \rangle \end{pmatrix} =
\begin{pmatrix} \frac{1}{2} & \pm\frac{\Delta}{2E_k} \\ \pm\frac{\Delta}{2E_k} &\frac{1}{2} \end{pmatrix},
\end{equation}
where the sign of the off-diagonal term depends on the specific spin $\sigma$ ($\mathcal{C}_{k\uparrow} = -\frac{\Delta}{2E_k}$ and $\mathcal{C}_{k\downarrow} =\frac{\Delta}{2E_k}$). After solving the characteristic equation $\det(\Gamma^{(q)}_{k, k+\pi, \sigma} - \lambda \I) = 0$, we obtain the eigenvalues of this matrix, which are strictly identical for both spin flavors:
\begin{align} \label{lambda_form}
    \lambda_k^{\pm} = \frac{1}{2} \pm \frac{\Delta}{2E_k}.
\end{align}
These two eigenvalues correspond to the two orthogonal eigenmodes that completely span the $(k, k+\pi)$ invariant subspace for a given spin $\sigma$. Since $\lambda_k^+ + \lambda_k^- = 1$ and the binary entropy function satisfies $\mathcal{F}(x) = \mathcal{F}(1-x)$, the mode-resolved entropy contribution $S_{k, k+\pi, \sigma}$ from this specific decoupled block is simply the sum of the binary entropies of these two independent eigenmodes [cf.~\cref{lambda_form}]:
\begin{equation}
    S_{k, k+\pi, \sigma} = \mathcal{F}(\lambda_k^+) + \mathcal{F}(\lambda_k^-) = 2\mathcal{F}(\lambda_k^+).
\end{equation}
Note that this contribution is identical for both spin flavors. Integrating $S_{k, k+\pi, \sigma}$ over the Brillouin zone $[-\pi, \pi]$ and summing over the spin flavors yields the total steady-state entanglement entropy $\se$, as the spin degeneracy trivially cancels the $1/2$ double-counting factor of the momentum pairs. In the thermodynamic limit $L \to \infty$, the discrete sum transitions into an integral over continuous momenta:
\begin{align} \label{eq:entropy_integral}
   \se (l)= \frac{1}{2} \sum_{k} \sum_{\sigma} S_{k, k+\pi, \sigma} = \sum_k 2\mathcal{F}(\lambda_k^+) = (L/2)\int_{-\pi}^{\pi} \frac{\d k}{2\pi} 2\mathcal{F}\left(\frac{1}{2} + \frac{\Delta}{2E_k}\right).
\end{align}
For the entanglement entropy $\se(l)$ on subsystem size $l=L/2$, this linear scaling with respect to the system size $L$ transparently proves the volume-law behavior:
\begin{equation}
    \se(l) = c_\Delta\cdot l=c_\Delta\cdot(L/2),
\end{equation}
where the entanglement entropy density $c_{\Delta}$ is explicitly defined as [cf.~Eq.(3) in the main text]:
\begin{equation}
    c_{\Delta} = 2\int_{-\pi}^{\pi} \frac{\d k}{2\pi} \mathcal{F}\left(\frac{1}{2} + \frac{\Delta}{2E_k}\right).
\end{equation}
As shown in Fig. 2(a,b) of the main text, $c_{\Delta}$ decreases monotonically with increasing $\Delta$, confirming that pairing suppresses the spatial propagation and delocalization of quasiparticles generated by the quench, thereby reducing the extensive entanglement in the steady state.

\subsection{Entanglement Saturation Time}
The entanglement growth following the quench can be intuitively understood through the semiclassical quasiparticle picture. The initial state acts as a uniform source of highly entangled Bogoliubov quasiparticle pairs. Once created, these entangled partners propagate ballistically in opposite directions across the lattice. The bipartite entanglement entropy between two spatial regions increases proportionally to the number of shared pairs where one quasiparticle has entered the first region and its partner has entered the second.

The propagation speed of a quasiparticle with momentum $k$ is dictated by its group velocity:
\begin{align}\label{eq:v_g}
    v_g(k) = \frac{\partial E_k}{\partial k} = -\frac{2J^2\sin(2k)}{\sqrt{(2J\cos k)^2+\Delta^2}}.
\end{align}
The time scale required for the entanglement to grow and eventually saturate to its volume-law steady state is bounded by the fastest propagating quasiparticles. For these entangled pairs to distribute entanglement across the entire system, they must propagate across the location of the half-chain bipartition, and the maximum travel distance from the boundary to the center is $L/2$. Therefore, the entanglement saturation time $\tau_\Delta$ is:
\begin{align}\label{eq:tau_L}
    \frac{\tau_\Delta}{L} = \frac{1}{2v_{\max}},
\end{align}
where $v_{\max} = \max_k |v_g(k)|$ denotes the maximum possible group velocity.

A larger pairing amplitude $\Delta$ flattens the quasiparticle dispersion band ($E_k \to \Delta$), which heavily suppresses the group velocities. Consequently, the quasiparticles propagate much more slowly, leading to a longer saturation time $\tau_\Delta$, as shown in Fig.~2(a,c) of the main text.

\bibliography{ref,library}